\newcommand{\bea}{\begin{eqnarray}}
\newcommand{\eea}{\end{eqnarray}}
\newcommand{\nn}{\nonumber}
\documentclass[%
 reprint,
 amsmath,amssymb,
 aps,
 prd,
 showkeys
]{revtex4-1}

\usepackage[USenglish]{babel}
\usepackage{graphicx}
\usepackage{dcolumn}
\usepackage{bm}
\usepackage{hyperref}

\begin{document}

\preprint{APS/123-QED}

\title{Thermo-magnetic evolution of the QCD strong coupling}

\author{Alejandro Ayala$^{1,2}$, C. A. Dominguez$^2$, Saul Hernandez-Ortiz$^1$,\\ L. A. Hernandez$^{1,2}$, M. Loewe$^{2,3,4}$, D. Manreza Paret$^{1,5}$ and R. Zamora$^{6,7}$}
\affiliation{%
$^1$Instituto de Ciencias Nucleares, Universidad Nacional Aut\'onoma de M\'exico, Apartado Postal 70-543, CdMx 04510, M\'exico.\\
$^2$Centre for Theoretical and Mathematical Physics, and Department of Physics, University of Cape Town, Rondebosch 7700, South Africa.\\
$^3$Instituto de F\'isica, Pontificia Universidad Cat\'olica de Chile, Casilla 306, Santiago 22, Chile.\\
$^4$Centro Cient\'ifico-Tecnol\'ogico de Valpara\'iso, Casilla 110-V, Valpara\'iso, Chile.\\
$^5$Facultad de F\'isica, Universidad de La Habana, San Lazaro y L, La Habana, Cuba.\\
$^6$Instituto de Ciencias B\'asicas, Universidad Diego Portales, Casilla 298-V, Santiago, Chile.\\
$^7$Centro de Investigaci\'on y Desarrollo en Ciencias Aeroespaciales (CIDCA), Fuerza A\'erea de Chile,  Santiago, Chile.
}%


\begin{abstract}
We study the one-loop gluon polarization tensor at zero and finite temperature in the presence of a magnetic field, to extract the thermo-magnetic evolution of the strong coupling $\alpha_s$. We analyze four distinct regimes, to wit, the small and large field cases, both at zero and at high temperature. From a renormalization group analysis we show that at zero temperature, either for small or large magnetic fields, and for a fixed transferred momentum $Q^2$, $\alpha_s$ grows with the field strength with respect to its vacuum value. However, at high temperature and also for a fixed value of $Q^2$ we find two different cases: When the magnetic field is even larger than the squared temperature, $\alpha_s$ also grows with the field strength. On the contrary, when the squared temperature is larger than the magnetic field, a turnover behavior occurs and $\alpha_s$ decreases with the field strength. This thermo-magnetic behavior of $\alpha_s$ can help explain the inverse magnetic catalysis phenomenon found by lattice QCD calculations. 
\end{abstract}

\pacs{Valid PACS appear here}
\keywords{Perturbative QCD, Magnetic fields, Finite temperature, Renormalization Group Equation}
\maketitle


Strongly interacting matter exhibits unusual properties when subject to magnetic fields. It has been shown by lattice QCD (LQCD) that both, the pseudo-critical temperature for the chiral or deconfinement phase transition, and the quark condensate for temperatures above the pseudo-critical temperature, decrease with the field strength. This phenomenon has been named \lq\lq inverse magnetic catalysis\rq\rq\ (IMC). The name implies the unexpected, and opposite behavior to the zero temperature case, whereby the quark condensate increases monotonically with the field strength~\cite{Bali}. Within a framework born out from LQCD simulations, this phenomenon can be attributed to the competition between the so called \lq\lq sea\rq\rq\ and \lq\lq valence\rq\rq\ contributions to the quark condensate, around the transition temperature. Indeed, when the condensate is computed from the QCD partition function, there are two distinct magnetic field-dependent factors: the determinant of the Dirac operator appearing when integrating out the fermion fields and the trace of the Dirac operator. The sea and valence contributions refer to the case where the magnetic field effect is only considered either in the determinant or in the trace of the Dirac operator, respectively. Although this separation seems artificial, it could apparently be placed on firmer grounds by resorting to LQCD techniques~\cite{Bruckmann}.

Another scenario to explain the origin of IMC is that the strong coupling receives thermo-magnetic corrections which make it increase or decrease depending on the competition between thermal and magnetic effects. The growth or decrease of the quark condensate would in turn be linked to the corresponding behavior of the coupling at zero or at high temperature, respectively. This scenario has been studied within effective QCD models~\cite{Farias,Ferreira,Ayala1,Ayala2,Ayala3,Avancini,Ayala4}, from the Schwinger-Dyson approach~\cite{Mueller}, and from the thermo-magnetic behavior of the quark-gluon vertex in QCD~\cite{vertex1,vertex2}. In the latter, it has been shown that the growth or decrease of the effective QCD coupling, at finite temperature and magnetic field strength, comes from a subtle competition between the color charges of gluons and quarks in such a way that at zero temperature, the former is larger than the latter whereas at high temperature, the coupling receives contributions only from the color charge associated to quarks.

In this work we study the thermo-magnetic behavior of the strong coupling $\alpha_s$. We resort to using the renormalization group equation (RGE), applied to the one-loop gluon vacuum polarization, computed in the presence of a magnetic field at zero and finite temperature. To understand the thermo-magnetic evolution of $\alpha_s$, we study the limiting regimes where the magnetic field is considered either in the weak or in the strong field case, both at zero and at high temperature. A renormalization group flow analysis for large ratios of the magnetic field to the temperature squared, applied to the relevant two fermion operator, has been recently carried out using the gauge-gravity correspondence in Ref.~\cite{Fito}. Also, a RGE analysis of the strong coupling running with temperature at zero magnetic field, based on the use of temperature dependent renormalization constants, has been carried out in Ref.~\cite{Steffens}

Let $P(q^2;T,|eB|;\alpha_s)$ be the un-renormalized coefficient of a given tensor structure upon which the gluon polarization can be decomposed at finite temperature $T$ and in the presence of a constant magnetic field $|eB|$. Since the thermo-magnetic medium breaks Lorentz invariance, $P$ can depend separately on the square of the components of the four-momentum $q^\mu$. However, for notational simplicity and latter convenience, we write $q^2$ for the momentum dependence. We scale the energy variables appearing in $P$ by the renormalization ultraviolet energy scale $\mu$, writing
\bea
   |eB|&=&\mu^2(|eB|/\mu^2),\,\,\,
   T^2=\mu^2(T^2/\mu^2),\nn\\
   q^2&=&\mu^2(q^2/\mu^2).
\label{differentmus}
\eea
Therefore, we have
\bea
   \!\!\!\!\!\!P(q^2;T,|eB|;\alpha_s)=\mu^D P(q^2/\mu^2;T^2/\mu^2,|eB|/\mu^2;\alpha_s),
\label{rescalemu}
\eea
where $D=2$ is the energy dimension of $P$. Since $\mu$ is arbitrary, the statement that $P$ should be independent of this scale is provided by the RGE~\cite{Pascual}, namely
\bea
   \!\!\!\!\left(\mu\frac{\partial}{\partial\mu}+\alpha_s\beta(\alpha_s)
   \frac{\partial}{\partial\alpha_s}-\gamma\right)
   P(q^2;T,eB;\alpha_s)=0,
   \label{RGEPi}
\eea
where $\beta(\alpha_s)$ is the QCD beta function defined by
\bea
   \alpha_s\beta(\alpha_s)=\mu\frac{\partial\alpha_s}{\partial\mu},
\eea
and 
\bea
   \gamma=\mu\frac{\partial}{\partial\mu}\ln Z^{-1},
\label{andim}
\eea
with $Z$ being the gluon vacuum polarization wave function renormalization. The beta function encodes the dependence of the strong coupling on the renormalization scale and thus the ultraviolet properties of the theory. It is well known that the QCD beta function is negative and that to one-loop level it is given by
\bea
   \beta(\alpha_s)&=&-b_1\alpha_s,\,\,\,\,
   b_1=\frac{1}{12\pi}\left(11N_c-2N_f\right),
   \label{beta1loop}
\eea
with $N_c$ the number of colors and $N_f$ the number of active flavors. 

To set up the stage for the analysis, let us first recall how the usual evolution of the strong coupling with the momentum scale is established. Starting from Eq.~(\ref{RGEPi}) and considering that the only energy scale in the function $P$ is $q^2$, we have,
\bea
   \!\!\!\!\!\!\!\!\!\left(\mu\frac{\partial}{\partial\mu}+\alpha_s\beta(\alpha_s)
   \frac{\partial}{\partial\alpha_s}-\gamma\right)
   P(q^2;\alpha_s)=0.
   \label{onlyq}
\eea
We now introduce the variable
\bea
   t=\ln(Q^2/\mu^2),
\label{changevar}
\eea
where $Q^2$ is the momentum transfered in a given process. Notice that the reference scale $\mu^2$ is usually large enough, so as to make sure that the calculation is well within the perturbative domain, therefore $Q^2 < \mu^2$. After this change of variable, the RGE becomes
\bea
   \left(-\frac{\partial}{\partial t}+\alpha_s\beta(\alpha_s)
   \frac{\partial}{\partial\alpha_s}-\gamma\right)P(q^2;\alpha_s)=0.
\label{afterchange}
\eea
Using the method of the characteristics~\cite{Fritz}, one obtains the relation between the coupling values evaluated at $Q^2$ and the reference scale $\mu^2$ as
\bea
   \int_{t(Q^2=\mu^2)}^{t(Q^2)}dt=-\frac{1}{b_1}\int_{\alpha_s(Q^2=\mu^2)}^{\alpha_s(Q^2)}
   \frac{d\alpha_s}{\alpha_s^2}.
\label{relation}
\eea
Solving for $\alpha_s(Q^2)$, we obtain
\bea
   \alpha_s(Q^2)=\frac{\alpha_s(\mu^2)}{1+b_1\alpha_s(\mu^2)\ln(Q^2/\mu^2)},
\label{solvong}
\eea
from where it is seen that as $Q^2$ increases, the coupling decreases.


In the presence of a magnetized medium and/or a heat bath, there appear extra energy scales in the analysis. The question we set up to answer is how the coupling evolves as a function of the magnetic field and/or temperature scales when we take as the reference scale the momentum transfered in the given processes, $Q^2$ and the smallest of either $T^2$ and $|eB|$. The strategy we follow is to use the explicit dependence of a certain $P$ on the corresponding energy variables to study the relationship that changes on these scales bare with changes on the coupling constant, by means of the RGE. We now proceed to the analysis for the case when the corresponding function $P$ is computed in the four limiting cases: the weak and strong magnetic field both at zero and high temperature.

In the weak field limit, namely $|eB| < Q^2$ and at $T=0$, the gluon vacuum polarization (omitting the diagonal color structure) can be expressed as~\cite{Hattori,nosotros} (see also Ref.~\cite{Aritra})
\bea
 \Pi_{\mbox{\tiny{weak}}}^{\mu \nu}=P^\parallel_{\mbox{\tiny{weak}}} \Pi_{\parallel}^{\mu\nu}
 +P^\perp_{\mbox{\tiny{weak}}}\Pi_{\perp}^{\mu\nu}
 + P^0_{\mbox{\tiny{weak}}}\Pi_{0}^{\mu\nu},
 \label{pol}
\eea
with
$\Pi_{\parallel}^{\mu\nu} = g^{\mu\nu}_\parallel-
  \frac{q^\mu_\parallel q^\nu_\parallel}{q^2_\parallel}$,
  $\Pi_{\perp}^{\mu\nu}  = g^{\mu\nu}_\perp+\frac{q^\mu_\perp q^\nu_\perp}{q^2_\perp}$  and
  $\Pi_{0}^{\mu\nu}  = g^{\mu\nu}-\frac{q^\mu q^\nu}{q^2} - \Pi_{\parallel}^{\mu\nu} -
  \Pi_{\perp}^{\mu\nu}$,
where, when taking the magnetic field along the $\hat{z}$ axis, we have $g^{\mu\nu}_\parallel=(1,0,0,-1)$, $g^{\mu\nu}_\perp=(0,-1,-1,0)$, $g^{\mu\nu}=g^{\mu\nu}_\parallel+g^{\mu\nu}_\perp$, $q^\mu_\parallel=g^{\mu\nu}_\parallel q_\nu$, $q^\mu_\perp=g^{\mu\nu}_\perp q_\nu$, $q^2_\parallel=(q_0)^2-(q_3)^2$, $q^2_\perp=(q_1)^2+(q_2)^2$ and $q^2=q^2_\parallel-q^2_\perp$. Let us look at the coefficient $P^\parallel_{\mbox{\tiny{weak}}}$. In Ref.~\cite{nosotros} it has been shown that with $N_f=3$ active flavors, its explicit expression is
\bea
   P^\parallel_{\mbox{\tiny{weak}}}\!\!&=&\!\!-\frac{2\alpha_s}{9\pi}|eB|^2
   \left[\frac{q_\parallel^2}{(q^2)^3}(2q_\parallel^2+q_\perp^2)\right],\nn\\
   \!\!&=&\!\!-\mu^2\frac{2\alpha_s}{9\pi}(|eB|^2/\mu^4)\nn\\
   \!\!&\times&\!\!
   \left[\frac{q_\parallel^2/\mu^2}{(q^2/\mu^2)^3}(2q_\parallel^2/\mu^2+q_\perp^2/\mu^2)\right],\nn\\
   \!\!&=&\!\!-\mu^2\frac{\lambda_B^4}{\lambda_q^2}\frac{2\alpha_s}{9\pi}(|eB|^2/\mu^4)\nn\\
   \!\!&\times&\!\!
   \left[\frac{q_\parallel^2/\mu^2}{(q^2/\mu^2)^3}(2q_\parallel^2/\mu^2+q_\perp^2/\mu^2)\right],\nonumber\\
\label{piparaweakB}
\eea
where in the second line we have scaled the momentum components and the square root of the magnetic field intensity by $\mu$ and in the third line we have introduced the dimensionless scale factors $\lambda_B$ and $\lambda_q$, one for each of the magnetic $\sqrt{|eB|}/\mu$ and momentum $q/\mu$ powers of the dimensionless ratios, respectively. Notice that Eq.~(\ref{piparaweakB}) satisfies
\bea
   \Big(\mu\frac{\partial}{\partial\mu} + \lambda_q\frac{\partial}{\partial\lambda_q} +
   \lambda_B\frac{\partial}{\partial\lambda_B} - D\Big)P^\parallel_{\mbox{\tiny{weak}}}=0.
\label{RGEPi1}
\eea
Using the RGE, Eq.~(\ref{onlyq}), we get
\bea
   \!\!\!\!\!\!\!\!\!\!\Big(\!\!-\lambda_q\frac{\partial}{\partial\lambda_q}
   -\lambda_B\frac{\partial}{\partial\lambda_B} 
   + \alpha_s\beta(\alpha_s)
   \frac{\partial}{\partial\alpha_s} -\tilde{\gamma}\Big)P^\parallel_{\mbox{\tiny{weak}}}=0,
\label{RGEPi11}
\eea
where $\tilde{\gamma}=\gamma-D$. Using the method of the characteristics, we can write
\bea
   dt=-\frac{d\lambda_q}{\lambda_q},\,\,\,\,\,\,
   dt=-\frac{d\lambda_B}{\lambda_B},
\label{weakfield1}
\eea
whose solutions are
\bea
   \lambda_q=C_qe^{-t},\,\,\,\,\,\,
   \lambda_B=C_Be^{-t},
\label{solsWB}
\eea
where $C_q$ and $C_B$ are integration constants to be determined from the initial condition for the evolution.
Upon combining Eqs.~(\ref{solsWB}), we can write
\bea
   \lambda_q+\lambda_B&=&(C_q+C_B)e^{-t}=e^{-t},
\label{combine}
\eea
where we have chosen that at $t=0$, for the the initial condition of the evolution, $\lambda_q=1$ and $\lambda_B=0$. Therefore $C_q+C_B=1$. Also, for the subsequent evolution we take $Q^2$ fixed and thus we refer the evolution of $|eB|$ to the reference scale $Q^2$, namely, we take $\lambda_B=|eB|/Q^2$. Therefore, we can write
\bea
   t=\ln\left(\frac{Q^2}{Q^2+|eB|}\right).
\label{eqt}
\eea
Notice that in this case, as opposed to Eq.~(\ref{changevar}), the evolution energy scale appears in the denominator of the logarithmic function in Eq.~(\ref{eqt}). Therefore, Eq.~(\ref{RGEPi11}) becomes
\bea
   \left(\frac{\partial}{\partial t}+\alpha_s\beta(\alpha_s)
   \frac{\partial}{\partial\alpha_s}-\tilde{\gamma}\right)P^\parallel_{\mbox{\tiny{weak}}}=0,
\label{afterchange2}
\eea
from where the relation between the coupling values evaluated at $|eB|$ and the reference scale $Q^2$ can be expressed as
\bea
   \int_{t(\lambda_B=0)}^{t(\lambda_B=|eB|/Q^2)}dt=-\frac{1}{b_1}\int_{\alpha_s(Q^2)}^{\alpha_s(Q^2+|eB|)}
   \frac{d\alpha_s}{\alpha_s^2}.
\label{relation}
\eea
Solving for $\alpha_s(Q^2+|eB|)$, we obtain
\bea
   \alpha_s(Q^2+|eB|)=\frac{\alpha_s(Q^2)}{1+b_1\alpha_s(Q^2)\ln\left(\frac{Q^2}{Q^2+|eB|}\right)}.
\label{solvong}
\eea
From Eq.~(\ref{solvong}), we see that as the magnetic field intensity increases, the coupling increases with respect to its corresponding value at the reference scale $Q^2$.

Next, let us consider the strong field limit $|eB| > Q^2$ still for $T=0$. Working in the Lowest Landau Level (LLL) approximation, the only non-vanishing coefficient is that of the structure $\Pi^{\mu\nu}_\parallel$, which for three active flavors and a vanishing quark mass ($m$) is given by~\cite{nosotros,Gusynin,Fukushima}
\bea
   \!\!\!\!\!\!\!\!\!P^\parallel_{\mbox{\tiny{strong}}}&=&-\frac{4}{3}\frac{\alpha_s}{\pi}|eB|
   e^{-q_\perp^2/2|eB|}\nn\\
      \!\!\!\!\!\!\!\!\!&=&-\mu^2\lambda_B^2\frac{4}{3}\frac{\alpha_s}{\pi}\left(\frac{|eB|}{\mu^2}\right)
   e^{-\left(\frac{\lambda_q^2}{\lambda_B^2}\right)(q_\perp^2/2|eB|)},
\label{piparastrongB}
\eea
where once again, in the second line we have scaled the momentum components and the square root of the magnetic field intensity by $\mu$ and have introduced the dimensionless scale factors $\lambda_B$ and $\lambda_q$, one for each of the magnetic $\sqrt{|eB|}/\mu$ and momentum $q/\mu$ powers of the dimensionless ratios, respectively. It is easy to check that  Eq.~(\ref{piparastrongB}) satisfies Eq.~(\ref{RGEPi1}) and thus, upon using the RGE, Eq.~(\ref{onlyq}), we obtain
\bea
   \!\!\!\!\!\!\!\!\!\!\Big(\!\!-\lambda_q\frac{\partial}{\partial\lambda_q}
   -\lambda_B\frac{\partial}{\partial\lambda_B} 
   + \alpha_s\beta(\alpha_s)
   \frac{\partial}{\partial\alpha_s} -\tilde{\gamma}\Big)P^\parallel_{\mbox{\tiny{strong}}}=0.
\label{RGEPi22}
\eea
Using the same arguments as for the weak field case, which immplies starting the evolution from the fixed scale $Q^2<|eB|$, we once again obtain for the relation between the coupling evaluated at $|eB|$ and the reference scale $Q^2$
\bea
   \alpha_s(|eB|)=\frac{\alpha_s(Q^2)}{1+b_1\alpha_s(Q^2)\ln\left(\frac{Q^2}{Q^2+|eB|}\right)}.
\label{solvongstrong}
\eea

The results in Eqs.~(\ref{solvong}) and~(\ref{solvongstrong}) show that for $T=0$, $\alpha_s$ is an increasing function of $|eB|$, when referred to the scale $Q^2$. This result, although not necessarily evident, is in retrospective almost trivial: Since the reference scale is taken as the square of the transfered momentum in a given process, the magnetic field scale $|eB|$ does not compete with any other energy scale and is thus the only one that drives the evolution. The reference scale $Q^2$, although large, is always smaller  than either the combined scale $Q^2 + |eB|$ or $|eB|$, which in turn determines the sign of the logarithmic function appearing either in Eq.~(\ref{solvong}) or~(\ref{solvongstrong}).

We now turn to study the finite temperature case. Given that there is no need to assume a given hierarchy between $T^2$ and $|eB|$, the calculation is more straightforwardly performed when working in the LLL approximation. This is a suitable approximation when one or the other scale is considered as being the largest one. As is shown in Ref.~\cite{nosotros}, in this approximation the coefficient for the parallel unrenormalized gluon polarization tensor structure is given by 
\bea
   P^\parallel_{\mbox{\tiny{T-B}}}
   &=&\mu^2\lambda_B^2\frac{4}{3}\frac{\alpha_s}
   {\pi}\left(\frac{|eB|}{\mu^2}\right)
   e^{-\left(\frac{\lambda_q^2}{\lambda_B^2}\right)(q_\perp^2/2|eB|)}
   \nn\\
   &\times&
   \ln \Big( \frac{\lambda_m^2 m^2/\mu^2}{\pi^2\lambda_T^2 T^2/\mu^2}
   \Big)\Bigg[ \frac{(q_3^2/\mu^2)+(q_0^2/\mu^2)}{(q_\parallel^2/\mu^2)} \Bigg],
\label{pilargeTweakB}
\eea
where we have again scaled the momentum components, the square root of the magnetic field, the quark mass and the temperature by $\mu$ and have introduced the dimensionless scale factors $\lambda_B$, $\lambda_q$, $\lambda_m$ and $\lambda_T$, one for each of the magnetic $\sqrt{|eB|}/\mu$, momentum $q/\mu$, mass $m/\mu$ and temperature $T/\mu$ powers of the dimensionless ratios, respectively. It is easy to check that Eq.~(\ref{pilargeTweakB}) satisfies
\bea
   \left(\mu\frac{\partial}{\partial\mu}\,\,+\!\!\sum_{i=q,B,m,T}
   \lambda_i\frac{\partial}{\partial\lambda_i}
   -D
   \right)
   P^\parallel_{\mbox{\tiny{T-B}}}=0
   \label{RGEPi3}
\eea
and that upon using the RGE, Eq.~(\ref{RGEPi}), one obtains
\bea
   \!\!\!\!\!\!\left(-\!\!\!\sum_{i=q,B,m,T}
   \lambda_i\frac{\partial}{\partial\lambda_i} + \alpha_s\beta(\alpha_s)
   \frac{\partial}{\partial\alpha_s} -\tilde{\gamma}\right)P^\parallel_{\mbox{\tiny{T-B}}}=0.
\label{RGEPi33}
\eea
As before, we can write
\bea
   \!\!\!\!\!\!\!\!\!\!\!dt=-\frac{d\lambda_q}{\lambda_q},\,\,
   dt=-\frac{d\lambda_B}{\lambda_B},\,\,
   dt=-\frac{d\lambda_m}{\lambda_m},\,\,
   dt=-\frac{d\lambda_T}{\lambda_T},
\label{weakfield2}
\eea
whose solutions are
\bea
   \lambda_q&=&C_qe^{-t},\,\,
   \lambda_B=C_Be^{-t},\,\,\,
   \lambda_m=C_me^{-t},\,\,\,
   \lambda_T=C_Te^{-t}, \nonumber \\
\label{solsWB2}
\eea
where $C_q$, $C_B$, $C_m$ and $C_T$ are integration constants to be determined from the initial condition for the evolution. We can again combine Eqs.~(\ref{solsWB2}) and write
\bea
   \!\!\!\!\!\!\lambda_q+\lambda_B+\lambda_m+\lambda_T=
   (C_q+C_B+C_m+C_T)e^{-t}.
\label{combinehighTandB}
\eea
Let us first consider the case where $|eB|$ is the largest of the energy (squared) scales. The lesson we learned from the analysis at $T=0$ is that one should chose the reference scale as the sum of the rest of the energies (squared) other than the one that is evolving. Since for the case of three active quark flavors the quark mass is small with respect to the rest of the energy scales involved, we neglect the quark mass and use as the reference scale the combination $Q^2+T^2$. Therefore, in a similar fashion to the analysis of the $T=0$ case, we obtain
\bea
   \!\!\!\!\!\!\alpha_s(|eB|)=\frac{\alpha_s(Q^2+\widetilde{T}^2)}{1+b_1\alpha_s(Q^2+\widetilde{T}^2)\ln\left(\frac{Q^2+T^2}{Q^2+T^2+|eB|}\right)},
\label{alphalargeBTneq0}
\eea
where, since the analysis is only valid at leading order, we have taken $\widetilde{T}$ as a representative fixed value of $T$ in the energy domain of interest to compute the value of the coupling at the reference scale. From Eq.~(\ref{alphalargeBTneq0}) we see that the coupling {\it grows} with the field intensity. 

\begin{figure}[t!]
 \begin{center}
  \includegraphics[scale=0.4]{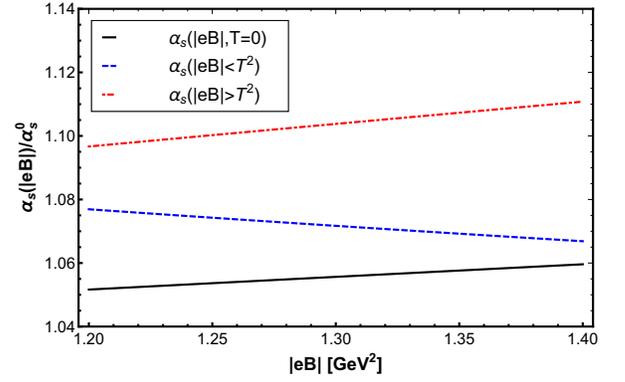}
 \end{center}
\caption{(Color on line) Strong coupling running as a function of the magnetic field strength. The curves are computed using a fixed transferred momentum $Q^2=1$ GeV$^2$. The black solid line shows the case for $T=0$. The blue dash line shows the case when $|eB|<T^2$ computed with $T=\widetilde{T}=1.5$ GeV. The red dash-dot line is the case for $|eB|>T^2$ computed with $T=1.1$ GeV and $|\widetilde{eB}|=1.2$ GeV$^2$.}\label{fig1}
\end{figure}
We now consider the case where $T^2$ is the largest of the energy (squared) scales. In this case the analysis yields
\bea
   \!\!\!\!\!\!\alpha_s(|eB|)=\frac{\alpha_s(Q^2+\widetilde{|eB|})}{1+b_1\alpha_s(Q^2+\widetilde{|eB|})\ln\left(\frac{Q^2+|eB|}{Q^2+|eB|+T^2}\right)},
\label{alphalargeTBneq0}
\eea
where we have also used a representative fixed value of the field strength $\widetilde{|eB|}$ to evaluate the coupling at the reference scale, to account for the fact that the analysis is only valid at leading order. From Eq.~(\ref{alphalargeTBneq0}) we see that the coupling {\it decreases} with the field intensity. We note that, contrary to the $T=0$ case, now the magnetic field has an extra scale to compete with. When the field intensity is the largest of these scales, the result is as for the $T=0$ case, where $|eB|$ is the only relevant reference scale and this in turn implies that the coupling increases. Conversely, when the temperature is the largest scale, the field strength appears both in the numerator and in the denominator of the argument of the logarithmic function, which in turn makes the coupling to decrease with the field strength.

Figure~\ref{fig1} shows the dependence of the strong coupling as a function of the magnetic field strength. We have chosen a fixed large value of $Q^2=1$ GeV$^2$. The black solid and red dash-dot lines show the cases for $T=0$ and $|eB|>T^2$, with $T=1.1$ GeV, respectively, corresponding to $|eB|$ being the largest of the energy scales. For both cases, we observe that the coupling increases with the field strength. The blue dash line shows the case for $|eB|<T^2$ with $T=1.5$ GeV. We observe that in this case, when the temperature is the largest of the energy scales, the coupling decreases as a function of the field strength. We also find the solution of the evolution equation for $P^\parallel$ in the four studied regimes. This is obtained in general as~\cite{Collins}
\bea
 P(\alpha_s)=P(\alpha_s^0)\exp\left[- \int_{\alpha_s^0}^{\alpha_s}d\alpha' \frac{\tilde{\gamma}(\alpha')}{\beta(\alpha')} \right],
\label{solRGE}
\eea
where, as before, $\tilde{\gamma}=\gamma-D$, with $\gamma$ the one-loop gluon field anomalous dimension~\cite{Grozin}, given explicitly by
\begin{equation}
 \gamma=\Bigg[ \Big( a-\frac{13}{3} \Big)C_A+\frac{8}{3}T_F N_f\Bigg]\frac{\alpha_s}{4\pi} \equiv A \ \alpha_s.
\end{equation}

We consider $N_f=3$ active flavors, work in the Landau gauge $a=0$, and remember that $T_F=1/2$ and $C_A=N_c$, with $N_c=3$ the number of colors. From Eq.~(\ref{solRGE}), we obtain that in each of the studied regimes, the parallel component of the gluon polarization tensor is given by
\bea
 \frac{P^\parallel_{\mbox{\tiny{weak,strong}}}(\alpha_s)}{P^\parallel_{\mbox{\tiny{weak,strong}}}(\alpha_s^0)}&=&\Bigg(\frac{Q^2+|eB|}{Q^2}\Bigg)^2\Bigg(\frac{\alpha_s(Q^2)}{\alpha_s(Q^2+|eB|)}\Bigg)^{\frac{A}{b_1}}, \nonumber \\
 \frac{P^\parallel_{\mbox{\tiny{$T<|eB|$}}}(\alpha_s)}{P^\parallel_{\mbox{\tiny{$T<|eB|$}}}(\alpha_s^0)}&=&\Bigg(\frac{Q^2+T^2+|eB|}{Q^2+\widetilde{T^2}}\Bigg)^2 \nonumber \\
 &\times&
\Bigg(\frac{\alpha_s(Q^2+\widetilde{T^2})}{\alpha_s(Q^2+T^2+|eB|)}\Bigg)^{\frac{A}{b_1}},
\nn\\
\frac{P^\parallel_{\mbox{\tiny{$T>|eB|$}}}(\alpha_s)}{P^\parallel_{\mbox{\tiny{$T>|eB|$}}}(\alpha_s^0)}&=&\Bigg(\frac{Q^2+T^2+|eB|}{Q^2+\widetilde{|eB|}}\Bigg)^2\nonumber \\
&\times& \Bigg(\frac{\alpha_s(Q^2+\widetilde{|eB|})}{\alpha_s(Q^2+T^2+|eB|)}\Bigg)^{\frac{A}{b_1}}.
\label{PilargeTsmallB}
\eea

We notice that the $P^\parallel$'s grow with the field strength. The growth is tamed by the temperature such that for the case with $T^2>|eB|$ the growth is less pronounced.

In conclusion, from an analysis of the RGE for the gluon polarization tensor in the presence of a magnetic field and at finite temperature, we have shown that when the magnetic field is the largest of the energy scales, the strong coupling increases with the field strength. Conversely, when the temperature is the largest of the scales, there is a turnover behavior and the strong coupling decreases as a function of the field strength. We emphasize that the renormalization program is performed integrating out the degrees of freedom above the ultraviolet scale $\mu$. Therefore, the running of the strong coupling accounts for the competition of magnetic and thermal energy scales much below $\mu$. This is in contrast to other schemes adopted for instance in Refs.~\cite{Steffens,Kapusta-Nakkagawa} where an attempt is made to integrate out degrees of freedom down to the thermal scale. Since the strength of the quark condensate is a measure of the quark-antiquark binding and this in turn depends on the strength of the QCD coupling, these results can help to understand the IMC phenomenon observed by LQCD.

This work was supported by Consejo Nacional de Ciencia y Tecnolog\'ia grant number 256494, by the National Research Foundaton (South Africa), by Fondecyt (Chile) grant numbers 1170107, 1150471, 11508427, Conicyt/PIA/Basal (Chile) grant number FB0821 and by CONICYT FONDECYT Iniciaci\'on under grant number 11160234. D. M. acknowledges support from a PAPIIT-DGAPA-UNAM fellowship.

\end{document}